# Relation between Dips of Prestack Time Migrated & Unmigrated Data

*Jagmeet Singh,*

*CEWELL, ONGC, Vadodara*

## Abstract

Relation between dips of post-stack migrated and unmigrated data is well known and easy to derive. A similar relation between dips of pre-stack migrated and unmigrated constant offset data is not available in literature and is calculated here. Since migration outputs for multiple offsets are available, the formula valid for all offsets gives us a tool to check the correctness of the migration velocity model. The formula is also used to study how a reflector with constant dip gets curved in constant offset gathers. Further, a handy formula for prestack migration aperture is derived and its variations with offset, depth and dip angle studied.

**Introduction**

Relation between dips of post-stack migrated and unmigrated data has been derived by Chun and Jacewitz (1981). A similar relation between dips of pre-stack migrated and unmigrated offset data is desirable, but not available in literature. We calculate the same here. Since migration outputs for different offsets are available, and the formula should hold for all these offsets, we get a tool to check the correctness of the migration velocity model. A handy formula for prestack migration aperture is also derived.

In Figure 1 below, we consider a dipping reflector with dip $\theta$. With a source placed at S and a receiver at R, we get reflection from point P' on the reflector, obtained by joining the virtual source S' with R by a straight line. The dotted curve is an ellipse corresponding to a fixed two way travel time S'P'R/v, where v is the velocity of the constant velocity medium.

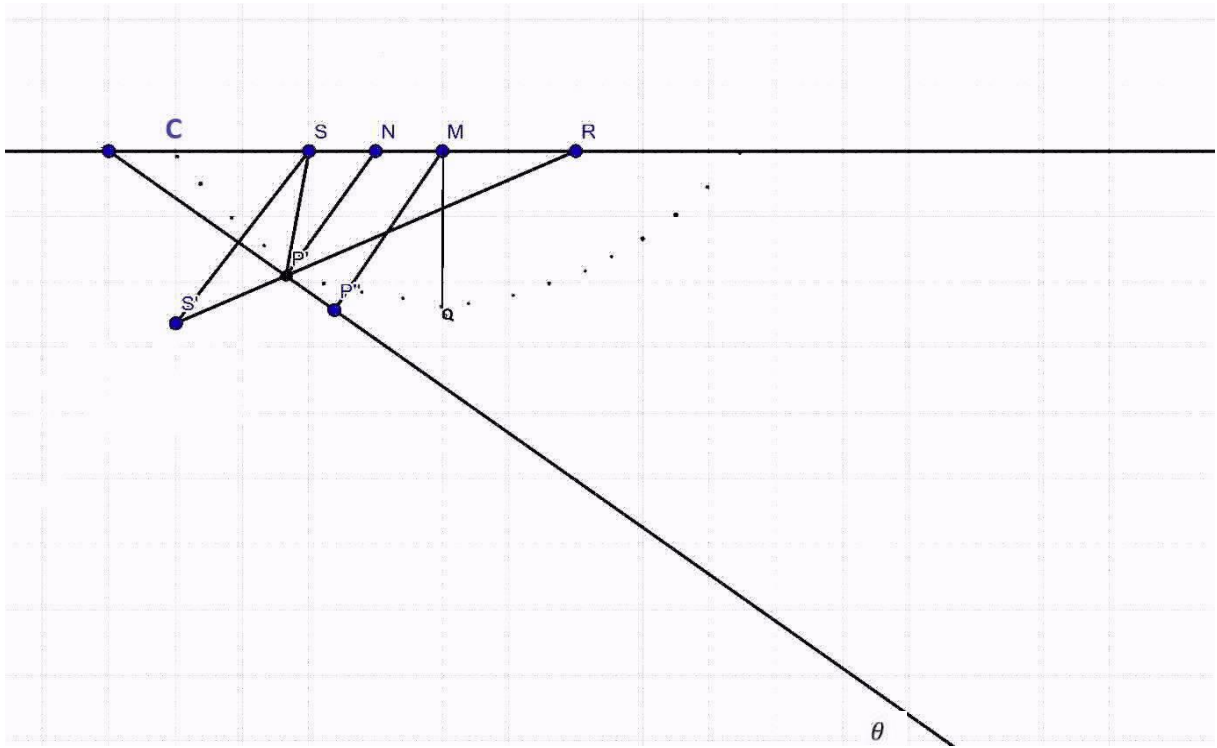

Taking SR=2h and SS'=d, we obtain from triangle SS'R:

$$(S'P'R)^2 = d^2 + 4h^2 + 4dh\sin\theta \qquad 1$$

Dividing both sides by $v^2$, an expression for two way time S'P'R/v can be obtained. Shifting the source S by $\delta x$ to the left or right (keeping offset unchanged) would also shift the point Q directly below midpoint M. Calculating the vertical and horizontal shifts should give us a formula for the apparent dip in terms of the real (i.e. prestack migrated) dip for a constant offset section. But before we proceed, we should understand that MQ/v does not equal two way time S'P'R/v. In fact S'P'R/v = SQR/v. Imagine a pencil at P' (with a taut thread SP'R) being moved around to form an ellipse so that when the pencil is at Q, the two way time equals SQR.

Two way time SQR/v > MQ/v, and would be plotted below Q on a time section (constant offset). Vertical shift in its position corresponding to a horizontal shift $\delta x$ can be calculated by differentiating (1) as follows:-

2(S'P'R) δ(S'P'R) = 2d δd + 4h δd sin θ

Or

$$(S'P'R)\, \delta(S'P'R) = \delta d\, (d + 2h \sin\theta) \qquad 2$$

It is easy to find a relationship between δd and δx :

$$\delta d = 2\, \delta x \sin\theta \qquad 3$$

Substituting (3) in (2), we get

$$\frac{\delta(S'P'R)}{\delta x} = \frac{2\sin\theta(d + 2h\sin\theta)}{\sqrt{d^2 + 4h^2 + 4dh\sin\theta}} \qquad 4$$

or

$$\tan\alpha_u = \frac{2\sin\theta(d + 2h\sin\theta)}{\sqrt{d^2 + 4h^2 + 4dh\sin\theta}}, \qquad 5$$

where $\alpha_u$ is the dip of the event on unmigrated offset section in (2 way)depth, and $\theta$ is the actual dip involving one way depth. Note that for h=0, (5) reduces to the well-known relation between unmigrated and migrated dips for the post-stack case:

$$\tan\alpha = \sin\theta, \qquad 6$$

where $\alpha$ is the dip on unmigrated offset section in depth i.e.

$$\tan\alpha = \frac{v\,\delta t}{2\delta x}, \qquad 7$$

Where $\delta t$ is the change in time (2 way) of the unmigrated event upon a change $\delta x$ in the lateral position x. Similarly, we write slope of the migrated event $\tan\theta$ as:

$$\tan\theta = \frac{v\,\delta\tau}{2\delta x}. \qquad 8$$

Substituting (7) and (8) in equation (6), we obtain

$$\frac{\delta\tau}{\delta x} = \frac{\delta t}{\delta x} \cdot \frac{1}{\sqrt{1 - \frac{v^2}{4}\left(\frac{\delta t}{\delta x}\right)^2}} \qquad 9$$

The above equation (9) is the same as that derived by Chun and Jacewitz in a different manner. Using (7) and (8) in equation (5), we obtain an extension of Chun and Jacewitz's formula for the prestack case:

$$\frac{\delta t}{\delta x} = \frac{\left(\dfrac{\frac{\delta\tau}{\delta x}}{\sqrt{1 + \frac{v^2}{4}\left(\frac{\delta\tau}{\delta x}\right)^2}}\right)\left(d + \dfrac{\frac{hv\delta\tau}{\delta x}}{\sqrt{1 + \frac{v^2}{4}\left(\frac{\delta\tau}{\delta x}\right)^2}}\right)}{\sqrt{d^2 + 4h^2 + 4dh\dfrac{\frac{v\,\delta\tau}{2\,\delta x}}{\sqrt{1 + \frac{v^2}{4}\left(\frac{\delta\tau}{\delta x}\right)^2}}}}. \qquad 10$$

Equation (10) gives a relationship between unmigrated dip (on a constant offset section) and the migrated dip, as we desired. It can be used to infer what values of dips to expect on different offset sections for a given value of the migrated dip $\frac{\delta\tau}{\delta x}$. If the migrated outcomes do not fit the formula, we would need to revise the velocity (or conclude that the 'event' is coherent noise). It is easy to check that for h=0, we get back the post-stack result i.e. equation (9). In equation (10), one would like to replace $d$, which is unknown on a seismic section, with a measurable and more meaningful quantity—we address this below.

From Figure (1), we see that $MP'' = \frac{d}{2} + hsin\theta$, and from Levin(1971) $2MP'' = vt_0$, where $t_0$ is the two-way zero offset time at midpoimt M. Using the above in (10). We get

$$\frac{\delta t}{\delta x} = \frac{vt_0 \frac{\delta \tau}{\delta x} \sqrt{\left(1 + \frac{v^2}{4}\left(\frac{\delta \tau}{\delta x}\right)^2\right)}}{\sqrt{v^2 t_0^2 + \frac{4h^2}{1 + \frac{v^2}{4}\left(\frac{\delta \tau}{\delta x}\right)^2}}} \quad (11)$$

Factoring out $vt_0$ from the denominator, we get

$$\frac{\delta t}{\delta x} = \frac{\frac{\delta \tau}{\delta x}}{\sqrt{1 + \frac{v^2}{4}\left(\frac{\delta \tau}{\delta x}\right)^2 + \frac{4h^2}{v^2 t_0^2}}} \quad (12)$$

Formula (12) was tested on synthetic data with different dips and offsets and found to be quite accurate. As a test case, in Figure 2, we show the results for a synthetic gather with offset=1350m, v=2km/s—measured $\frac{\delta t}{\delta x} = 0.2 ms/m$, $\frac{\delta \tau}{\delta x} = 0.22 \ ms/m$, $t_0 = 1.725s$, whereas predicted $\frac{\delta t}{\delta x}$ from the above equation comes out to be 0.2007 ms/m, an error of just 0.35%. Equation (12) also brings out the important role played by offset-depth ratio $\frac{2h}{vt_0}$ in seismic applications—importance of such a role was also e.g. brought out in Singh (2020).

Next, we see that for a given offset, changing d leads to dips varying with time on constant offset sections i.e. a planar reflector gets curved on a constant offset section. In Figure 3 below, we plot unmigrated angle $\alpha$ vs. d for a reflector with a true dip $\theta$ of 30 degrees for offsets=400 m, 800 m and 1600 m as shown. We see that the reflector gets curved or has different dips at different depths/times, with the dip converging to a constant value at high depths.

In Figure 4, we plot angle $\alpha$ vs. dip for a reflector with true dip $\theta$ = 60 degrees. We see a faster convergence to a constant value. In Figure 5 below, the three curves for different offsets as above have merged into a straight line at a constant value of $\alpha$ = 45 degrees for $\theta$ = 90 degrees, as is expected from formula (5) above.

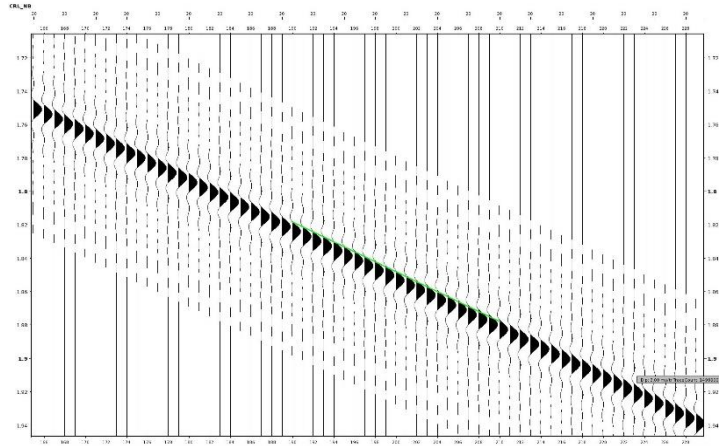

(a) Constant offset gather—dip measured to be 3ms/tr (trace interval = 15m). Offset=1350m.

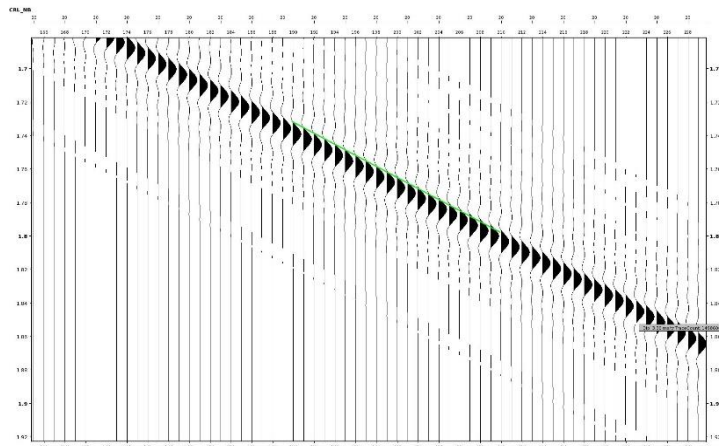

(b) Migrated Constant offset gather—dip measured to be 3.3 ms/tr

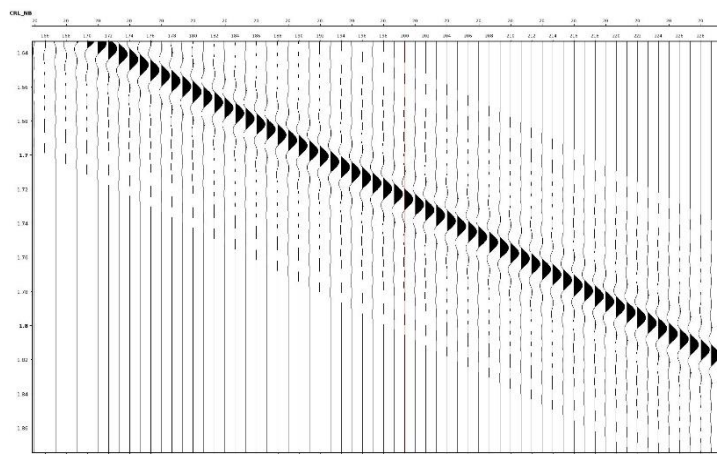

(c) Zero offset time $t_0$ measured to be 1.725s on NMO'ed constant offset gather for the central trace around which dip was measured

Figure 2

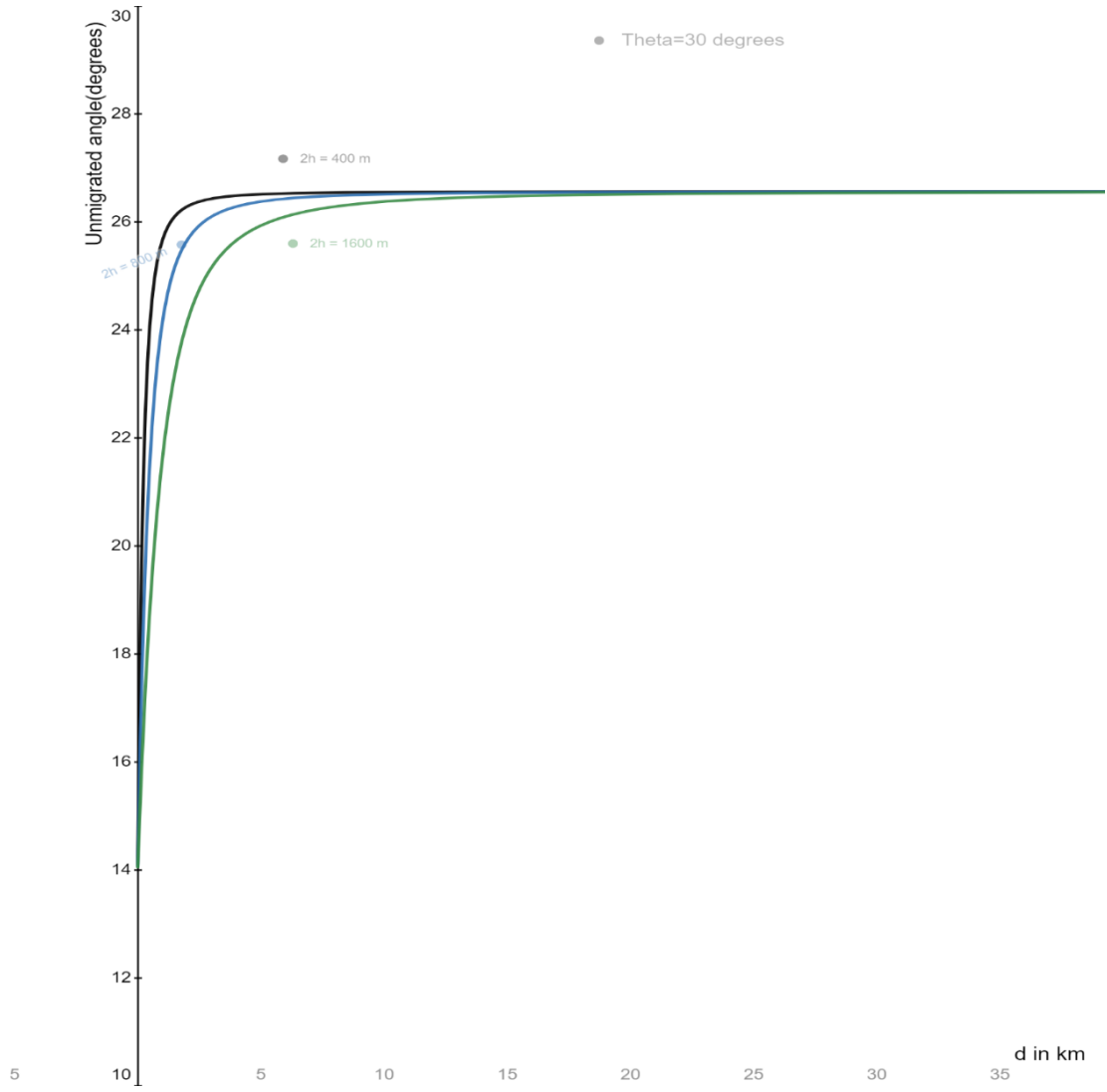

Figure 3

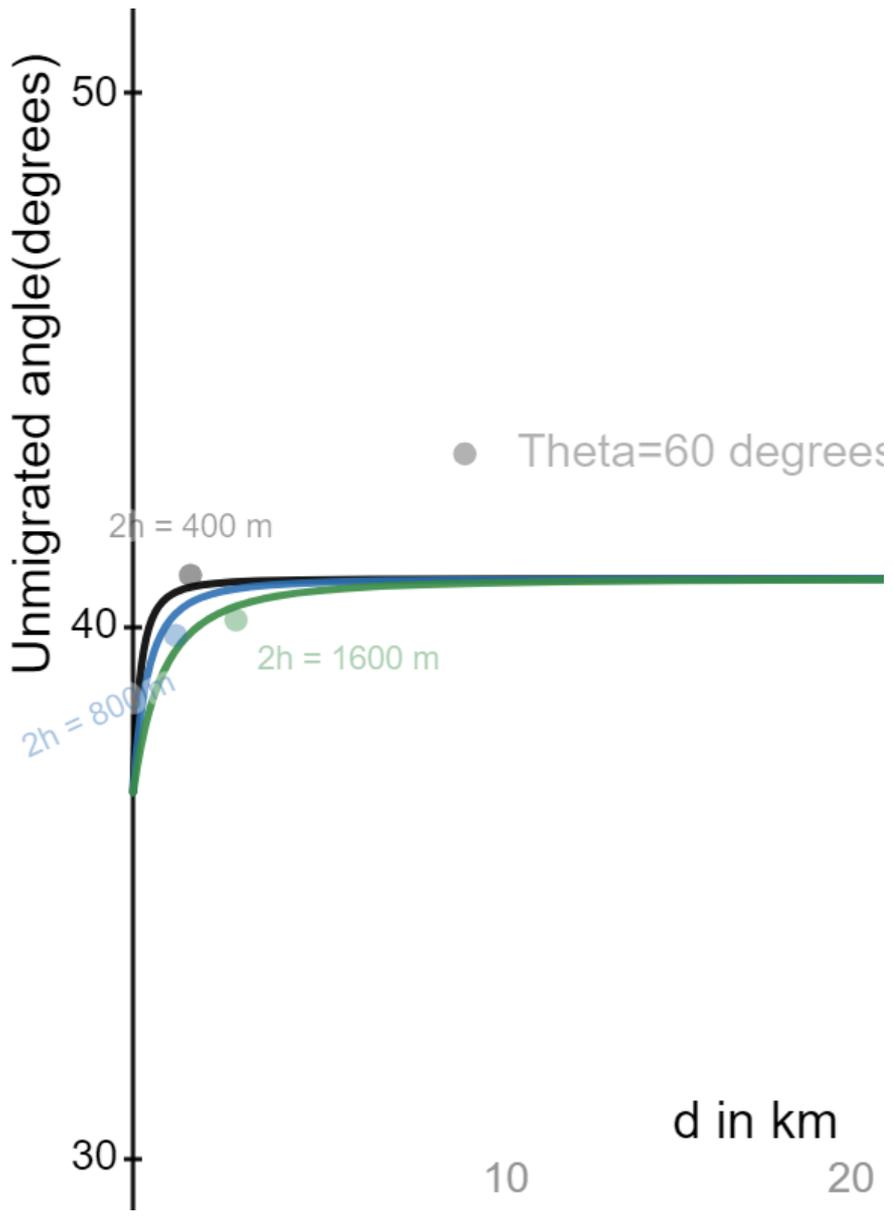
Figure 4

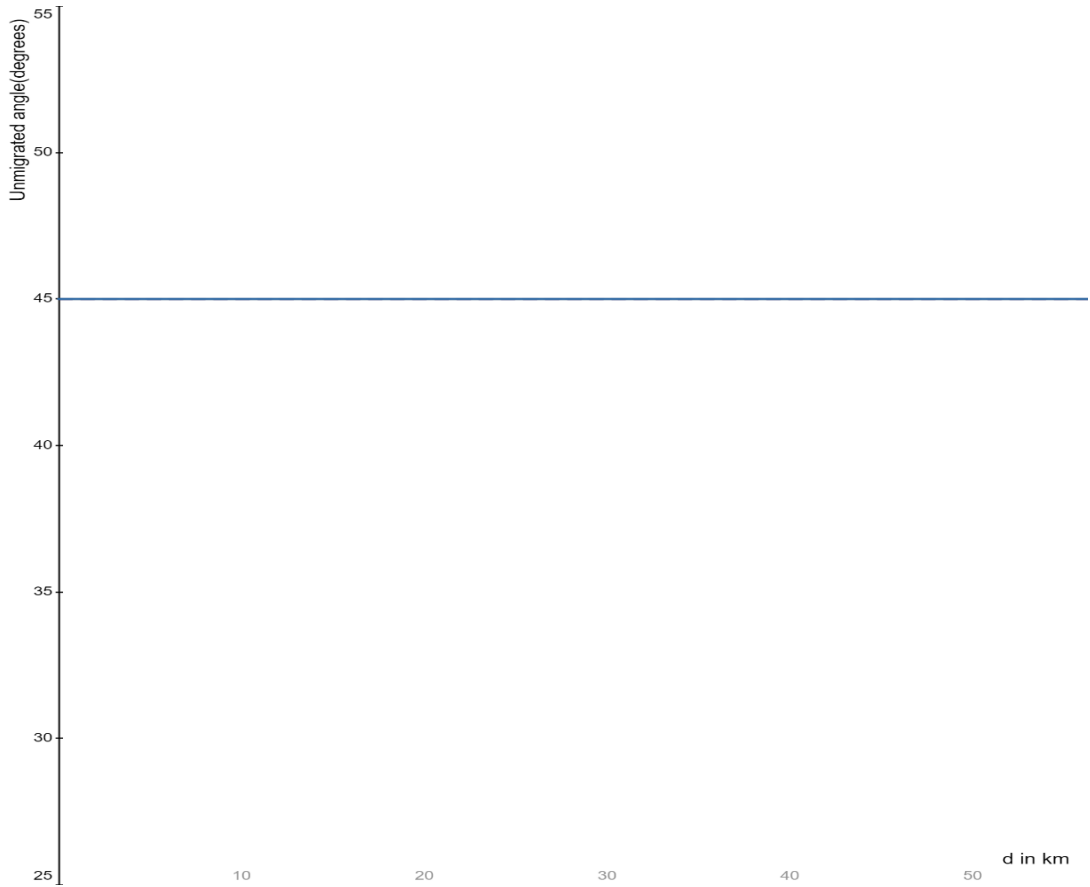

Figure 5

We now turn our attention to deriving a formula for prestack migration aperture, another area of interest, as it would allow us to optimize choice of aperture width for different offsets and depths. Referring back to figure 1, we see that we can write an equation for the ellipse if we know the semi-major axes i.e. MC and MQ. Clearly $\frac{MC}{v} = \frac{t}{2}$, where t is the two way time, and as discussed above MQ/v is not the two way time, but zero offset time corresponding to ray path SQR for the case of zero dip. So

$$\frac{MQ}{v} = t_0' = \sqrt{t^2 - \frac{4h^2}{v^2}} \qquad 13$$

Note that we have used the expression $t_0'$ for the two way time MQ/v to distinguish it from the two way time $t_0 = \frac{MP''}{v}$ corresponding to the dipping reflector. Using the above results for MC and MQ, we get the following equation for the migration ellipse:-

$$\frac{x^2}{v^2 t^2} + \frac{z^2}{v^2 t_0'^2} = 1 \qquad 14$$

Differentiating (14), we get

$$\frac{2x\,dx}{v^2 t^2} + \frac{2z\,dz}{v^2 t_0'^2} = 0$$

i.e.
$$\frac{dz}{dx} = -\frac{xt_0'^2}{zt^2}$$

or
$$x = z\,tan\theta\,\frac{t^2}{t^2-4h^2/v^2}, \qquad 15$$

where we have replaced dz/dx by $tan\,\theta$, the reflector slope(as the reflector is tangential to the ellipse at P') and the zero offset zero dip time $t_0'$ has been expressed in terms of two way time $t$. A similar equation where we have distances 'a' and 'b' in place of $t$ and $t_0'$ has been derived by Murty and Shankar (2006). But equation (15) involving two way times is much more intuitive and calculation friendly.

Let us re-write equation (15) in a manner more useful for graphical analysis. Say point (x,z) corresponds to point P' in Figure 1. It is easy to see that $z = d\,cos\theta$. Further using equation (1) in equation(15), we re-write equation (15) as

$$x = d\,sin\theta\left(\frac{d^2 + 4h^2 + 4dh\,sin\theta}{d^2 + 4dh\,sin\theta}\right) \qquad 16$$

We study the variation of full migration aperture 2x with full offset 2h in Figure 6 below (for $\theta = 30$ degrees). Three different values of d i.e. 1 km, 2 km and 10 km have been taken. We see that aperture is more sensitive to offset than d (related to depth) ---a rule of thumb seems to be that full aperture (2x) equals full offset or source-receiver distance, provided offset to depth ratio is high.

In Figure 7, we show the variation of aperture 2x with dip angle for 2h =2 km and d =2 km. As expected, aperture increases with dip angle for a given d and h.

### Conclusions

We have extended Chun and Jacewitz's formula relating migrated slope to unmigrated slope for the post stack to the pre stack case i.e. from zero offset to finite offset. Redundancy of data i.e. availability of data at multiple offsets gives us a tool to check the correctness of our rms velocity model. Migration aperture as a function of depth and offset is also derived, enabling a judicious choice of the same for running PSTM (prestack time migration).

### Acknowledements

The author expresses his sincere thanks to Shri Abhishek Kumar, SG(S), RCC, ONGC, Baroda for his help in verifying formula (12) on synthetic data.

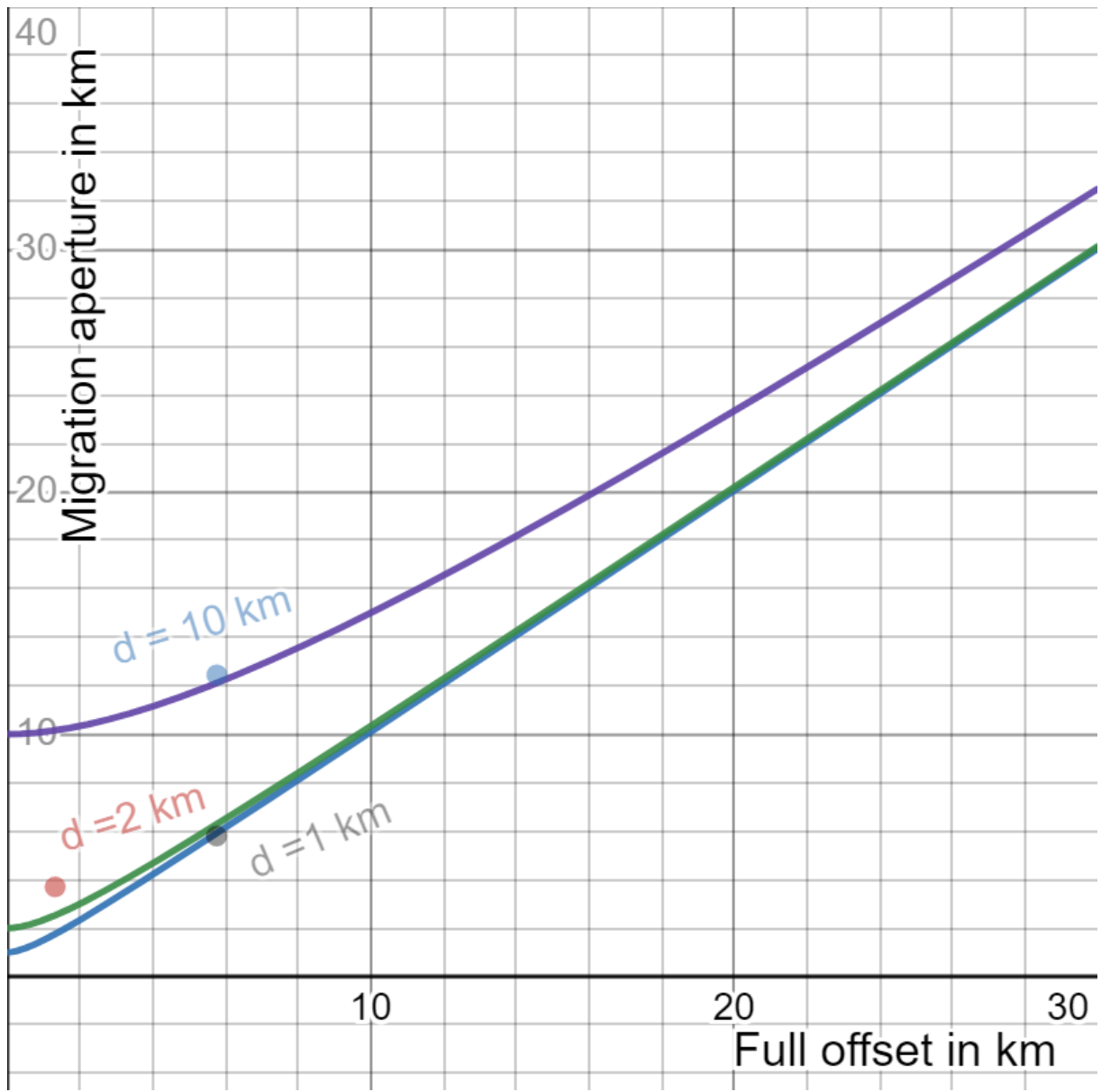

Figure 6

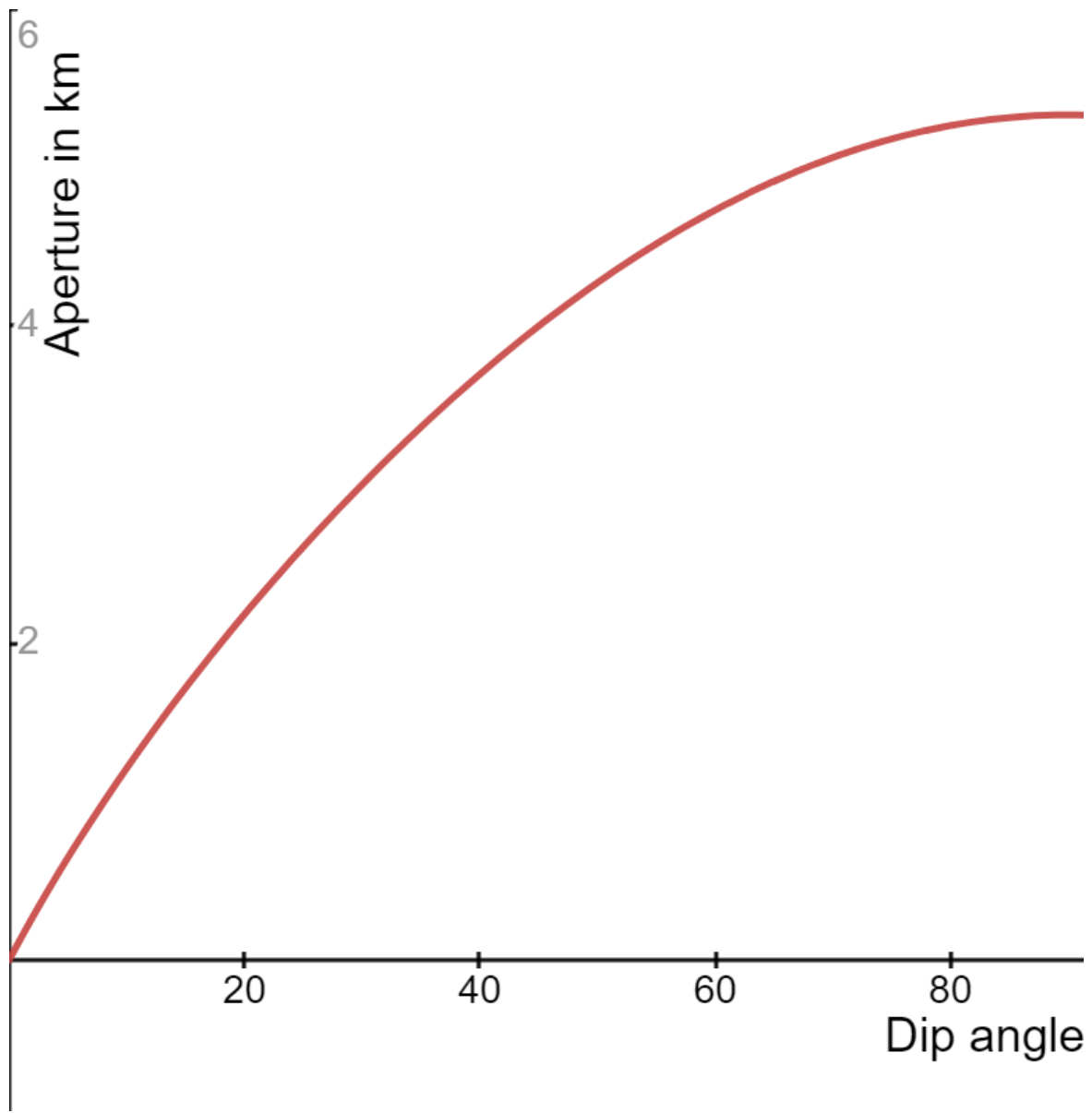

Figure 7

**References**

Chun, J.H. and Jacewitz, C., 1981, Fundamentals of frequency-domain migration: Geophysics, 46, 717–732.

Murty, J.V.S.S.N. and Shankar, T., 2006, Pre Stack Migration Aperture-An Overview, Geohorizons, Vol. 11, No.2, 8-13

Levin, Franklyn K.. 1971, "Apparent Velocity from Dipping Interface Reflections." *Geophysics* **36**. 510-516.

Singh Jagmeet, 2020, Velocity uncertainty and stack power analysis, 13th Biennial International Conference and Exhibition Kochi, Society of Petroleum Geophysicists, India